%% file: conference_101719.tex
\documentclass[conference]{IEEEtran}
\IEEEoverridecommandlockouts
\usepackage{cite}
\usepackage{bm}
\usepackage{float}
\usepackage{amsmath,amssymb,amsfonts}
\usepackage{mathtools, nccmath}
\usepackage{graphicx}
\usepackage{textcomp}
\usepackage{xcolor}
\usepackage{caption}
\usepackage{subcaption}
\usepackage{nicematrix}
\usepackage[shortlabels,inline]{enumitem}
\usepackage{comment}
\usepackage{cuted}
\usepackage[para,flushleft]{threeparttable}
\usepackage{adjustbox}
\usepackage{booktabs}
\usepackage{color, colortbl}
\usepackage{soul}
\usepackage{multirow}
\usepackage{tikz}
\usepackage{pgfplots}
\usepackage{algorithm}
\usepackage{siunitx}
\usepackage[noend]{algpseudocode}%
\usepackage[hidelinks]{hyperref}
\usepackage{arydshln}
\usepackage{svg}

\definecolor{Gray}{gray}{0.9}
\def\BibTeX{{\rm B\kern-.05em{\sc i\kern-.025em b}\kern-.08em
    T\kern-.1667em\lower.7ex\hbox{E}\kern-.125emX}}

\makeatletter
\newcommand{\linebreakand}{%
  \end{@IEEEauthorhalign}
  \hfill\mbox{}\par
  \mbox{}\hfill\begin{@IEEEauthorhalign}
}
\makeatother

\begin{document}

\title{fpgaHART: A toolflow for throughput-oriented acceleration of 3D CNNs for HAR onto FPGAs}

\author{
    \IEEEauthorblockN{Petros Toupas$^{1,2}$}
    \and
    \IEEEauthorblockN{Christos-Savvas Bouganis$^1$}
    \and
    \IEEEauthorblockN{Dimitrios Tzovaras$^2$}
    \linebreakand
    \IEEEauthorblockA{
    $^1$
    Dpt. of Electrical and Electronic Engineering\\
    Imperial College London\\
    Email: \{p.toupas21,christos-savvas.bouganis\}@imperial.ac.uk}
    \and
    \IEEEauthorblockA{
    $^2$
    Information Technologies Institute\\
    Centre of Research and Technology Hellas\\
    Email: \{ptoupas,dimitrios.tzovaras\}@iti.gr}
}

\maketitle

\begin{abstract}
Surveillance systems, autonomous vehicles, human monitoring systems, and video retrieval are just few of the many applications in which 3D Convolutional Neural Networks are exploited. However, their extensive use is restricted by their high computational and memory requirements, especially when integrated into systems with limited resources. This study proposes a toolflow that optimises the mapping of 3D CNN models for Human Action Recognition onto FPGA devices, taking into account FPGA resources and off-chip memory characteristics. The proposed system employs Synchronous Dataflow (SDF) graphs to model the designs and introduces transformations to expand and explore the design space, resulting in high-throughput designs. A variety of 3D CNN models were evaluated using the proposed toolflow on multiple FPGA devices, demonstrating its potential to deliver competitive performance compared to earlier hand-tuned and model-specific designs.
\end{abstract}


\input{Introduction/introduction}

\input{Related_Work/related_work}

\input{Modeling/modeling}

\input{Design_Space_Exploration/dse}

\input{Evaluation/evaluation}

\input{Conclusions/conclusions}

\newpage

\bibliographystyle{IEEEtran}
\bibliography{IEEEabrv,references}

\end{document}

%% file: Introduction/introduction.tex
\section{Introduction} \label{intro}

Two-dimensional CNNs have excelled in image-related tasks in recent years. The increasing importance and amount of applications arising from video-related tasks, such as video surveillance, autonomous driving, and elderly monitoring, has demanded the development of algorithms that incorporate and account for the temporal domain. Three-dimensional CNNs are one of the most common approaches used to deal with video and volumetric data. With the addition of a new dimension, such as time or depth, 3D CNNs augment their capability to learn by extracting information related to the newly added dimension. 

\begin{figure}[]
    \centering
    \includegraphics[width=0.95\linewidth,keepaspectratio]{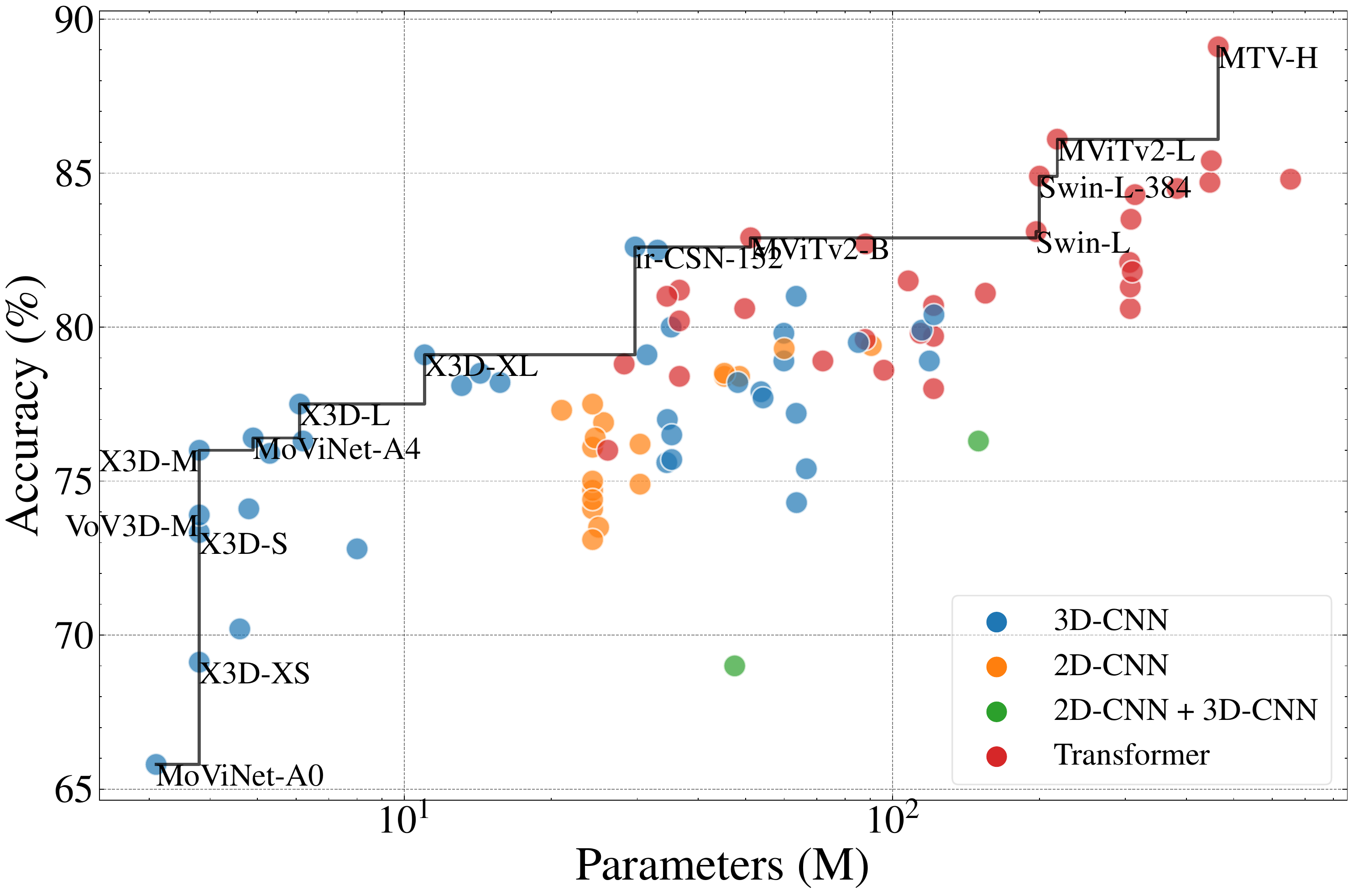}
    \caption{Kinetics-400 pareto is dominated by 3D-CNNs for small number of parameters. Demonstrating the deployability of 3D-CNNs on edge devices with limited resources.}
    \label{fig:pareto_over_years}
\end{figure}

3D CNNs have exhibited outstanding performance, particularly in the task of Human Action Recognition (HAR). The use of 3D CNNs allows the interpretation of human motion across video frames, allowing the detection of a wide range of human actions without the requirement for specific time domain approaches like LSTMs. As can be seen in Figure \ref{fig:pareto_over_years}, 3D CNNs dominate the pareto front in one of the most widely used HAR benchmarks, Kinetics-400, while the recent emergence of vision transformers has also begun to drive some designs to the pareto front, however such networks require orders of magnitude additional GFLOPs to operate.

While 3D CNNs are capable of capturing time or depth-related features, the additional dimension of the input frequently results in greater workloads, computational and memory requirements compared to 2D CNNs. Numerous hardware devices, including GPUs, FPGAs, and ASICs, have been used to mitigate for the 3D CNNs' high processing requirements and provide high performing systems. The current work aims to design systems that can be deployed to FPGA devices, due to their flexibility in adapting to the requirements of such evolving field as well as with their potential for achieving high performance and low power consumption.

In HAR, given a single input video clip, $N$ new clips are generated by shifting a (fixed) time window throughout the original clip's duration, and $M$ new clips are generated by cropping an area (for each image in the clip). The final evaluation of the original clip is acquired by passing each of the $N \times M$ generated clips through the HAR model and averaging their predictions. As such, upon deployment of such models, it is necessary to process the input video segment multiple times to maintain the desired performance. Therefore, throughput-oriented designs and solutions are of high interest. The key contributions of this paper are the following:
\begin{itemize}[noitemsep,topsep=0pt]
    \item Introduction of fpgaHART. A throughput-oriented toolflow for optimising and mapping 3D CNNs to FPGAs, supporting a variety of models and devices, while taking into account the model characteristics, available platform resources, and memory bandwidth characteristics.
    \item The expansion of the SDF graph model used for capturing performance requirements in CNN mapping to streaming architectures to explicitly handle irregular blocks with branching, which are commonly utilised in modern 3D CNN HAR models.
    \item A comprehensive evaluation, utilising various devices and models, including cutting-edge 3D CNN HAR models that have yet to be explored. The findings lay the groundwork for the computation of HAR models on FPGAs for throughput-oriented applications.
\end{itemize}

%% file: Related_Work/related_work.tex
\section{Background} \label{background}

Although 3D CNNs have been around for a while, there have only been a few papers aimed at their acceleration on FPGAs. The majority of these works focus on relatively old 3D CNNs, such as the C3D \cite{Ji2013} model, whose performance falls short of state-of-the-art models. Fan et al. introduced a series of works on 3D CNN acceleration for HAR on FPGA systems, \cite{Fan2017, Fan2018, Fan2019}. In their initial work \cite{Fan2017}, they proposed the F-C3D hardware architecture for the acceleration of C3D \cite{Ji2013}, which is capable of supporting multiple 3D convolutional layers and design strategies for overcoming the challenges associated with 3D CNNs while also allowing their design to be ported to other FPGA devices. In their subsequent work \cite{Fan2018}, they proposed an analytical model and a tool for optimising the hardware architecture based on the device specification and accuracy requirements, as well as the use of block floating point (BFP) arithmetic precision to minimise accuracy loss and the need for retraining the model. In their most recent work, \cite{Fan2019}, they proposed E3DNet, an efficient 3D CNN based on their proposed 3D-1 bottleneck building block. Their hardware implementation of E3DNet, named F-E3D, is capable of real-time performance at the execution time of 35.3 ms per clip\footnote{A clip is defined as a stacked sequence of frames that are meant to be the input of the 3D CNN}, while achieving an accuracy of 85.1\%\footnote{H. Duan et. al. \cite{Duan2020} currently holds the SoA results on UCF101 achieving 98.6\% accuracy} on the UCF101 benchmark.

Liu et al. \cite{Liu2019AFpgas} proposed a unified hardware architecture for 2D and 3D CNN acceleration based on the observation that the computing patterns of 2D and 3D CNNs are similar. They convert CNN convolutions to matrix multiplication operations, paying close attention to memory optimizations in order to overcome the difficulties of feature map replications. Additionally, they employed an analytical model to configure the accelerators for optimal resource use. They have targeted and evaluated their design on C3D model. Shen et al. \cite{Shen2018} followed a similar approach, developing a unified template-based architecture based on the Winograd algorithm capable of handling both 2D and 3D CNNs. Additionally, they developed an analytical technique for efficiently exploring the design space for mapping 2D and 3D CNNs on FPGA accelerators. The authors have targeted the C3D model for the evaluation of their proposed design. Sun et al. \cite{Sun20203DPruning} used a blockwise pruning approach to apply weight pruning to two distinct 3D CNN architectures, namely C3D and R(2+1)\cite{Tran2018}. Their hardware design, which is based on the Alternating Direction Method of Multipliers (ADMM), together with the suggested pruning approach, enables the acceleration of 3D CNNs with low accuracy loss compared to the unpruned version. Toupas et al. \cite{Toupas2023FMM-X3D:Recognition} recently proposed a throughput-oriented hardware design for X3D, a modern and state-of-the-art 3D CNN, with an emphasis on automating model branches management. Additionally, they have recently introduced a toolflow named HARFLOW3D \cite{ToupasHARFLOW3D:Devices} that simplifies the mapping and optimisation of 3D CNN models on FPGA devices, delivering promising results on latency-focused applications.

The majority of research has been focused on the C3D \cite{Ji2013} model for HAR, which was introduced in 2013. The model's architecture is rather simplistic, consisting of only sixteen consecutive layers, and it performs poorly in terms of accuracy when compared to the modern SoA models in HAR ($85.2$\% in UCF101 compared to $98.6$\% which is the current SoA). In terms of design complexity, it is comparable to the LeNet or AlexNet in the three-dimensional space. Due to the fact that the aforementioned approaches are essentially dedicated to the design of the target model, it is unclear how they may be extended, evaluated or perform in the more complicated architectures of modern state-of-the-art HAR models. This study focuses on supporting more recent 3D CNNs as well, which have a significantly larger number of layers and deviate from the sequential approach of early networks by containing branching within Resnet-like blocks.

%% file: Modeling/modeling.tex
\section{Hardware-Level Interpretation} \label{modeling}

This section discusses hardware-level 3D CNN model interpretation and modelling. The work is inspired by fpgaConvNet \cite{Venieris2019}, a framework that automatically maps 2D CNN models to FPGA platforms, and extends it in significant ways, as outlined below. The proposed framework extracts the parameters of each layer in a Directed Acyclic Graph (DAG) and the connections between layers from a high-level description of a 3D CNN model. The network's supported layers are mapped to parametrisable hardware building blocks that implement their functionality. Subsequently, the framework generates the network's Synchronous Data-Flow Graph (SDFG) by mapping the DAG nodes to their hardware equivalent blocks and adding them as nodes and arcs in the SDFG. Finally, using the SDF computation model, a network configuration's SDFG node performance is estimated. The sections below describe the proposed tool's components.

\subsection{3D CNN layers as DAG nodes}

The description of a neural network model supplied by high-level frameworks such as pytorch and onnx is comprised by three main parts. First, the layers and their connections that define the model's structure and flow. Second, each layer's special attributes and configuration, and finally the actual values of the learnable parameters associated with their layers (if any). A dedicated model parser is developed, parsing the above descriptions to build a DAG containing all of the relevant information of the neural network. The DAG structure is faithful to the original, retaining just the essential information from the layers' specific attributes and configuration. Additionally, the parser stores the model parameters/weights for future use during inference. Table \ref{tab:dag_symbols} summarises the symbols used to denote the parameters of DAG nodes that represent and characterise the layers of the models.

\begin{table}[h]
    \centering
    \caption{DAG nodes parameters symbols.}
    \label{tab:dag_symbols}
    \resizebox{\columnwidth}{!}{%
    \begin{tabular}{ll}
    \toprule
    \textbf{Symbols} & \textbf{Definitions} \\
    \midrule
        $\bm{Sz_{\text{\tt i}}}$ & size dimensions of the input feature map \\
        $\bm{Sz_{\text{\tt o}}}$ & size dimensions of the output feature map \\ 
        $K_{\text{\tt h}}, K_{\text{\tt w}}, K_{\text{\tt d}}$ & height, width and depth of convolution kernel \\ 
        $St_{\text{\tt h}}, St_{\text{\tt w}}, St_{\text{\tt d}}$ & stride value on height, width and depth dimensions \\ 
        $Pd_{\text{\tt h}}, Pd_{\text{\tt w}}, Pd_{\text{\tt d}}$ & padding value on height, width and depth dimensions \\ 
        $Gp$ & number of groups in which the input is split \\ 
        & along the channel axis on convolution layers \\ 
        $T$ & type of activation or element-wise function \\ 
        $M$ & mode of element-wise operation (normal/broadcasting) \\ 
    \bottomrule
    \end{tabular}
    }
\end{table}

The following data structures are utilised by the tool to capture the layers of the 3D CNN models:
\begin{itemize}[noitemsep,topsep=0pt]
    \item \textbf{3D Convolutional and Pooling Layers} 
        The following types of convolutional/pooling layers are supported: \begin{enumerate*}[(a)]
            \item spatial convolution/pooling $K_h \times K_w \times 1$,
            \item temporal convolution/pooling $1 \times 1 \times K_d$,
            \item depth-wise convolution/pooling,
            \item point-wise convolution/pooling
        \end{enumerate*}. The configuration of the layer as stored in a DAG node is as follows, $\bm{<Sz_i, Sz_o, K, St, Pd}, Gp>$,
        where:
        \begin{itemize}[label=$\circ$]
            \item \textbf{K} is a 3-value vector $[K_h, K_w, K_d]$ specifying the depth, height and width of the 3D conv window.
            \item \textbf{St} is a 3-value vector $[St_h, St_w, St_d]$ specifying the strides of the convolution along each dimension.
            \item \textbf{Pd} is a 3-value vector $[Pd_h, Pd_w, Pd_d]$ denoting the amount of padding applied to each dimension.
        \end{itemize}
    \item \textbf{3D Activation Layers} 
        The activation functions supported are the following: \begin{enumerate*}[(a)]
            \item ReLu activation,
            \item Sigmoid activation,
            \item Swish activation which is expressed as $ y = x * sigmoid(x) $:
        \end{enumerate*}, with its DAG's layer structure, $\bm{<Sz_i, Sz_o, T>}$.
    \item \textbf{3D Element-wise Layers} 
        Element-wise operations are layers that combine (add, mul) data from several branches. These layers combine several inputs into a single output, where the shapes of the inputs may or may not be identical, resulting in different functionality (normal vs broadcasting). The layers configuration as a DAG node, $\bm{<Sp_{i1}, ..., Sp_{iN}, Sz_o, T, M>}$.
    \item \textbf{3D Global Average Pooling Layer} 
    While the standard pooling operation samples patches of the input feature map to decrease its size, GAP samples the whole feature map into a single value, creating an output vector with the same shape as the channels. DAG's layer configuration: $\bm{<Sz_i, Sz_o>}$.
\end{itemize}

\subsection{SDFG representation with branch support}\label{sdfg_representation}

To take advantage of the SDF model's capabilities, the tool maps DAG nodes into their associated hardware building blocks, which implement the functionality of each layer in the underlying hardware. Using SDF theory, the SDFG may be represented as a topology matrix $\Gamma$. The nodes are represented by the columns of this matrix, while the arcs that link the nodes are represented by the rows. The data consumption/production rates for each node in each arc can be inferred by looking the element at $(node, arc)$ position in the $\Gamma$ matrix. Positive values, by convention, drive data production, whereas negative ones drive data consumption. The element $\Gamma(n,a)=-1$, for example, indicates that node n consumes data at arc \textit{a} at a rate of one.

The $\Gamma$ matrix is decomposed into several matrices (as show in Eq. \ref{gamma_matrix}), allowing a more in-depth examination of each one separately and more fine control overall. The initial decomposition of the $\Gamma$ matrix yielded three distinct matrices:
\begin{enumerate}[i)]
    \item The stream matrix $\textbf{S}$. This matrix element stores the number of incoming and outgoing parallel streams that arrive to each node's input and output.
    \item The rate matrix $\textbf{R}$. The rate matrix elements include the normalised data production and consumption rates of each node at each arc (number of elements produced/consumed per cycle). The values in this matrix range from 0 to 1.
    \item The data matrix $\textbf{C}$. The width of each individual stream from the $\textit{S}$ matrix is stored in this matrix elements. Since all of the streams are assumed to have the same bit width of 16, the above matrix is not taken into consideration in this study.
\end{enumerate}
\useshortskip
\begin{equation}\label{gamma_matrix}
    \Gamma = S \times R
\end{equation}

The upper bi-diagonal structure of the $\Gamma$ matrix prevents the modelling of branching behaviours, i.e. graphs with nodes receiving multiple incoming arcs and nodes with many outgoing arcs. This work proposes and implements modifications to the SDFG structure to ease the building of graphs with several incoming or outgoing arcs at nodes, hence supporting branching models without the need to explicitly define them with static predefined layers. The depth of each side of a branch is computed to incorporate some extra buffering for the streams that are combined at the merge points in order to ensure the flow of data across the design's streams as well as to equalise the rates at the merge points.

\subsection{3D CNN layers as hardware building blocks} \label{hw_building_blocks}

The hardware building blocks are the major components utilised to construct the SDFG, which will be used subsequently to estimate the network's performance. The configuration of these blocks, in conjunction with the network's topology, is utilised to automatically generate and construct the design's synthesisable Vitis HLS code. The representation of the supported hardware building blocks comprises of the following: 
    \begin{enumerate}[i),noitemsep,topsep=0pt]
        \item \textbf{DAG parameters}. A set of parameters that originated from the layer's settings as a DAG node. These settings are the layers' structural configuration that cannot be changed.
        \item \textbf{SDFG parameters}. An additional set of parameters which have an impact on the layer's performance and are the ones that the optimisation algorithm searches for during the design space exploration phase.
    \end{enumerate}

\begin{table}[h]
    \label{tab:symbols}
    \centering
    \caption{SDFG nodes parameters symbols.}
    \resizebox{0.9\columnwidth}{!}{%
    \begin{tabular}{ll}
    \toprule
    \textbf{Symbols} & \textbf{Definitions} \\
    \midrule
        $s_{\text{\tt i}}$ & number of streams at the layer's input channels \\
        $s_{\text{\tt o}}$ & number of streams at the layer's output filters \\
        $r_{\text{\tt i}}$ & consumption rate of the layer \\
        $r_{\text{\tt o}}$ & production rate of the layer \\
        $p_{\text{\tt mac}}$ & number of parallel multiply and accumulate (MAC) \\ 
        & operations in a convolution layer \\
    \bottomrule
    \end{tabular}
    }
\end{table}

The hardware building block representation of the 3D CNN layers is described below:
\begin{itemize}[noitemsep,topsep=0pt]
        \item \textbf{3D Convolutional/Pooling Layers:} %
            \[
                \begin{aligned}
                    \bm{<{DAG_{params}}, {s_{i}, s_{o}, r_{i}, r_{o}, p_{mac}}>}
                \end{aligned}
            \]
            The \textit{$s_{i}$},\textit{$s_{o}$}, and \textit{$p_{mac}$} are altered during the DSE and affect the final performance of the layer. Meanwhile the \textit{$r_{i}$} and \textit{$r_{o}$} depend on the \textit{$p_{mac}$} which means they are implicitly altered during the DSE as well. A mode detailed analysis of the convolution layer and its sub-modules is provided in fpgaConvNet \cite{Venieris2019}. 
        \item \textbf{3D Activation, 3D Global Average Pooling Layers:} %
            \[
                \begin{aligned}
                    \bm{<{DAG_{params}}, {s_{i}, s_{o}, r_{i}, r_{o}}>}
                \end{aligned}
            \] 
            These layers' \textit{$r_{i}$} and \textit{$r_{o}$} can achieve consumption/production rates of 1 (if not constraint from previous layers or the memory rates), due to their element-wise functionality and the simplicity of their operations. The only exception here is the 3D Global Average Pooling, in which $r_{o} = \frac{1}{H \times W \times D}$, where $H$ is the height, $W$ is the width, and $D$ is the depth dimension of the input feature map.
        \item \textbf{3D Element-wise Layers:} %
            \[
                \begin{aligned}
                    \bm{<{DAG_{params}}, {s_{i1}, s_{i2}, s_{o}, r_{i1}, r_{i2}, r_{o}}>}
                \end{aligned}
            \]
            This layer's \textit{$r_{i1}$}, \textit{$r_{i2}$} and \textit{$r_{o}$} can achieve consumption/production rates of 1 (if not constraint from previous layers or the memory rates), due to their element-wise functionality and the simplicity of their operations. It should be noted that in cases when the rates in either of the inputs are restricted owing to a lower production rate of a previous layer or due to memory constraints, the layer's input rates are equalised to the lower consumption rate among them.
\end{itemize}

%% file: Design_Space_Exploration/dse.tex
\section{Design Space Exploration} \label{dse}

The hardware mapping of the SDFG assumes a final streaming architecture to be inferred. Each design point in the design space has a specific combination of the involved layers' tunable parameters as they were described in section \ref{hw_building_blocks}. Essentially using a set of transformations operating on the SDFG, the aforementioned parameters are being altered while the design space is being explored by simulated annealing, the heuristic optimisation algorithm used in this study.

\subsection{3D CNN Model Partitioning}

CNN hardware architecture design incorporates two distinct approaches. Single computation engines implement a time-shared processing unit and a scheduler, while streaming architectures like the one presented employ a hardware block for each CNN layer to better exploit per layer parallelism. When trying to fit all the layers into a single design without reconfiguring the FPGA, the more layers a CNN has or the larger the input, the more FPGA resources are used, limiting each layer's parallelism.

Through utilising the FPGA's reconfiguration capabilities, network execution can be split into smaller partitions to solve this problem. By producing a unique architecture and delivering a bitstream for each partition, it allows the design of more finely tuned architectures that better fit each layer. This approach also drastically reduces off-chip memory access to only the design's input and output streams, allowing on-chip memory to be used for data reuse. This strategy requires reconfiguration every time a new partition is loaded, however increasing the batch size can amortise this cost.

Beginning with a random partitioning of the model's layers by introducing $L$ initial reconfiguration points, the optimisation process gradually modifies these partitions. The alterations to the partitions are focused on two key concepts:
\begin{itemize}
    \item The optimizer detects partitions that limit the performance of the model because they are memory constrained, have fully exploited the parallelism of their layers, or do not have sufficient resources to exploit enough parallelism from their layers.
    \item Out of the candidate partitions to be modified, the optimiser selects the partitions with the lowest performance and moves layers from or to adjacent partitions with a goal of improving their performance.
\end{itemize}
Between stages that modify existing partitions, the optimiser independently executes a series of partition-specific optimisation steps based on coarse and fine transformations as detailed below.

\subsection{Partition-Specific Optimisations}

Each partition layer's configurable parameters leverage its parallelism based on two factors:
\begin{itemize}
    \item \underline{The number of parallel executions of coarse operations} in each layer, which depends on the input feature map's channels. The primary operations of each layer can be performed in parallel by deploying multiple processing blocks up to the number of channels. 3D convolutional layers can exploit both input channels and output filters coarse-level parallelism. The $s_{i}$ and $s_{o}$ parameters of the hardware building block configuration are updated and searched for optimal values throughout the DSE to realise this parallelism. These variables affect the stream matrix $\textbf{S}$, which affects the topology matrix $\Gamma$, which determines design performance.
    
    \item \underline{The dot product operation's parallelism} during the kernel's convolution with a given input volume piece on 3D convolutional layers. This parallelism determines the number of parallel multipliers and the depth of the adder tree for additions. A completely unrolled design uses $N$ multipliers and $N-1$ adders, yielding 1 dot product per cycle, but restricting the setup to a single multiplier and adder yields $1/N$ dot products per cycle, where $N$ is the input and kernel shape. There is a trade-off between performance and resource utilisation. The DSE optimises the $p_{mac}$ parameters of the 3D convolutional layer hardware building block configuration to accomplish this parallelism. These variables affect the rate matrix $\textbf{R}$, which affects the topology matrix $\Gamma$, which estimates design performance.
\end{itemize}

\subsection{Performance Modelling}

To describe the performance of a given design based on the topology matrix $\Gamma$, an additional matrix reflecting the workload of each layer is included. As the topology matrix provides the throughput of each layer at its input in consumptions/cycle and output in productions/cycle, constructing a matrix with the total workload of each layer, i.e. the total number of elements to be consumed and produced,  allows the generation of a new matrix that provides the number of cycles each layer requires to consume its workload. More specifically a workload matrix $\textbf{W}$ has the same structure as the topology matrix $\Gamma$. By element-wise dividing the $\textbf{W}$ matrix with the $\Gamma$ matrix, the final $II$ matrix is being calculated as shown below:
\useshortskip
\begin{equation}\label{ii_matrix}
    II = W / \Gamma
\end{equation}
The $II$ is the initiation interval matrix, and its entries represent the total number of cycles required by each layer to consume its workload completely. The maximum value of the $II$ matrix, denoted by $II_{max}$ determines the initiation interval of the whole SDFG. The total execution time of a partition with batch size B is given by the following equation:
\useshortskip
\begin{equation}\label{exec_time}
    \operatorname{t(B,\Gamma)}=\frac{1}{\textrm{clock rate}}\cdot (D + II_{max}\cdot (B-1))
\end{equation}
where $D$ is the total number of cycles needed to fill the pipeline depth of the whole design, and its calculated by adding the depths of each layer and the depth added due to the extra buffering to deal with the branches in the design.

In order to capture the model's overall execution time, the execution times of each individual partition are summed up with the addition of the total reconfiguration time:
\useshortskip
\begin{equation}\label{exec_time_total}
    \operatorname{t_{total}(B,\Gamma)}=\sum_{n=0}^{N_p}t_n(B,\Gamma_i) + (N_p-1)\cdot t_{reconfig}
\end{equation}
where $N_p$ is the total number of the partitions of the model, and $t_{reconfig}$ is the reconfiguration time for loading a partition to the FPGA. As can be noticed from Eq. \ref{exec_time_total}, the extra overhead caused by the device reconfiguration is proportional to the number of partitions of the final solution and is independent of the batch size. By increasing the number of batches processed by the model, the first term dominates the execution time and the cost of reconfiguration is amortised.

Finally the overall throughput of the proposed architecture is inferred by dividing the total workload of the model in GOps (Giga Operations) times the batch size, with the total execution time:
\useshortskip
\begin{equation} \label{tota_throughput}
    \operatorname{Throughput(B)}=\frac{Workload_{model}*B}{t_{total}(B,\Gamma)}
\end{equation}

The design space exploration for each partition is described as an optimization problem with the following objective: $max(t(B,\Gamma)), s.t. rsc(\Gamma) \leq rsc_{avail}$. As this is a non-convex optimisation problem, its optimisation is based on the simulated annealing heuristic algorithm algorithm that attempts to maximise the design's throughput while ensuring that FPGA resource use does not exceed the available resources.

%% file: Evaluation/evaluation.tex
\section{Evaluation} \label{evaluation}

\begin{table}
\centering
\caption{3D CNN models characteristics}\label{tab:models_characteristics}
\resizebox{\columnwidth}{!}{%
\begin{threeparttable}
\begin{tabular}{cc:c:c:c:c}
\hline
                                & C3D       & Slowonly      & R(2+1)D-18        & R(2+1)D-34        & X3D \\
\hline
FLOPs (G)\tnote{$\dagger$}                       & 38.61     & 54.81         & 8.52              & 12.91             & 6.97  \\
Parameters (M)                  & 78.41     & 32.51         & 33.41             & 63.72             & 3.82  \\
Num. of Layers                  & 27        & 174           & 82                & 154               & 396   \\
Num. of Conv Layers             & 8         & 53            & 37                & 69                & 115   \\
Spatial dimensions              & $112 \times 112$   & $256 \times 256$    & $112 \times 112$    & $112 \times 112$    & $256 \times 256$    \\
Num. of Frames                  & 16        & 8             & 16                & 16                & 16    \\
\begin{tabular}[c]{@{}c@{}}UCF101\\ Accuracy (\%)\end{tabular}          & 83.2      & 94.54         & 88.66             & 92.27             & 96.52 \\
\hline
\end{tabular}
\begin{tablenotes}
    \item[$\dagger$] FLOPs are reported as MAC operations.
\end{tablenotes}
\end{threeparttable}
}
\end{table}

\begin{table*}[]
\caption{Comparison with existing works on 3D CNN HAR models}
\label{tab:performance_comparison}
\resizebox{\textwidth}{!}{%
\begin{threeparttable}
\begin{tabular}{@{}cccccccccccccc@{}}
\toprule
 & H. Fan{\cite{Fan2017}} & H. Fan{\cite{Fan2018}} & Z. Liu{\cite{Liu2019AFpgas}} & \multicolumn{2}{c}{J. Shen \cite{Shen2018}\tnote{$\ddagger$}} & \multicolumn{2}{c}{M. Sun\cite{Sun20203DPruning}} & H. Fan{\cite{Fan2019}} & \multicolumn{5}{c}{Ours} \\ \midrule \midrule

Model & \multicolumn{1}{c|}{C3D} & \multicolumn{1}{c|}{C3D} & \multicolumn{1}{c|}{C3D} & \multicolumn{2}{c|}{C3D} & C3D & \multicolumn{1}{c|}{R(2+)D-18} & \multicolumn{1}{c|}{E3D} & C3D & Slowonly & R(2+1)D-18 & R(2+1)D-34 & X3D \\

GFLOPs\tnote{$\ast$} & \multicolumn{1}{c|}{38.61} & \multicolumn{1}{c|}{38.61} & \multicolumn{1}{c|}{38.61} & \multicolumn{2}{c|}{-} & 38.61 & \multicolumn{1}{c|}{8.52} & \multicolumn{1}{c|}{6.1} & 38.61 & 54.9 & 8.52 & 12.91 & 6.97 \\

\rowcolor{lightgray} Accuracy (\%) & \multicolumn{1}{c|}{79.87} & \multicolumn{1}{c|}{81.99} & \multicolumn{1}{c|}{83.2} & \multicolumn{2}{c|}{83.2} & 83.2 & \multicolumn{1}{c|}{88.66} & \multicolumn{1}{c|}{85.17} & 83.2 & 94.54 & 88.66 & 92.27 & 96.52 \\

FPGA & \multicolumn{1}{c|}{ZC706} & \multicolumn{1}{c|}{ZC706} & \multicolumn{1}{c|}{VC709} & VC709 & \multicolumn{1}{c|}{VUS440} & ZCU102 & \multicolumn{1}{c|}{ZCU102} & \multicolumn{1}{c|}{Intel SX660} & ZCU102 & ZCU102 & ZCU102 & ZCU102 & ZCU102 \\

\rowcolor{lightgray} clips/s\tnote{$\dagger$} & \multicolumn{1}{c|}{1.84} & \multicolumn{1}{c|}{2.09} & \multicolumn{1}{c|}{8.65} & 11.18 & \multicolumn{1}{c|}{20.36} & 2.05 & \multicolumn{1}{c|}{4.11} & \multicolumn{1}{c|}{28.32} & 3.38 & 2.54 & 4.62 & 2.63 & 13.44 \\

GOps/s\tnote{$\dagger$} & \multicolumn{1}{c|}{70.41} & \multicolumn{1}{c|}{80.12} & \multicolumn{1}{c|}{330.74} & 427.29 & \multicolumn{1}{c|}{778} & 78.44 & \multicolumn{1}{c|}{111.71} & \multicolumn{1}{c|}{172.8} & 130.84 & 144.44 & 39.59 & 34.26 & 85.96 \\

GOps/s/DSP\tnote{$\dagger$} & \multicolumn{1}{c|}{0.087} & \multicolumn{1}{c|}{0.103} & \multicolumn{1}{c|}{0.092} & 0.281 & \multicolumn{1}{c|}{0.511} & 0.065 & \multicolumn{1}{c|}{0.092} & \multicolumn{1}{c|}{0.109} & 0.052 & 0.057 & 0.015 & 0.013 & 0.034 \\

Op/DSP/cycle\tnote{$\dagger$} & \multicolumn{1}{c|}{0.511} & \multicolumn{1}{c|}{0.519} & \multicolumn{1}{c|}{0.774} & 1.874 & \multicolumn{1}{c|}{2.559} & 0.435 & \multicolumn{1}{c|}{0.613} & \multicolumn{1}{c|}{0.727} & 0.325 & 0.358 & 0.098 & 0.084 & 0.213 \\

Frequency (MHz) & \multicolumn{1}{c|}{172} & \multicolumn{1}{c|}{200} & \multicolumn{1}{c|}{120} & 150 & \multicolumn{1}{c|}{200} & 150 & \multicolumn{1}{c|}{150} & \multicolumn{1}{c|}{150} & 160 & 160 & 160 & 160 & 160 \\

Precision & \multicolumn{1}{c|}{\begin{tabular}[c]{@{}c@{}}fp-16\end{tabular}} & \multicolumn{1}{c|}{BFP} & \multicolumn{1}{c|}{\begin{tabular}[c]{@{}c@{}}fp-16\end{tabular}} & \begin{tabular}[c]{@{}c@{}}fp-16\end{tabular} & \multicolumn{1}{c|}{\begin{tabular}[c]{@{}c@{}}fp-16\end{tabular}} & \begin{tabular}[c]{@{}c@{}}fp-16\end{tabular} & \multicolumn{1}{c|}{\begin{tabular}[c]{@{}c@{}}fp-16\end{tabular}} & \multicolumn{1}{c|}{\begin{tabular}[c]{@{}c@{}}float-32\end{tabular}} & \begin{tabular}[c]{@{}c@{}}fp-16\end{tabular} & \begin{tabular}[c]{@{}c@{}}fp-16\end{tabular} & \begin{tabular}[c]{@{}c@{}}fp-16\end{tabular} & \begin{tabular}[c]{@{}c@{}}fp-16\end{tabular} & \begin{tabular}[c]{@{}c@{}}fp-16\end{tabular} \\

DSP (\%) & \multicolumn{1}{c|}{90} & \multicolumn{1}{c|}{86.6} & \multicolumn{1}{c|}{99.8} & 42 & \multicolumn{1}{c|}{53} & 48 & \multicolumn{1}{c|}{48} & \multicolumn{1}{c|}{93.3} & 51.49 & 63.77 & 66.21 & 66.46 & 84.43 \\

BRAM (\%) & \multicolumn{1}{c|}{86.6} & \multicolumn{1}{c|}{88.1} & \multicolumn{1}{c|}{26.6} & 52 & \multicolumn{1}{c|}{30} & 100 & \multicolumn{1}{c|}{100} & \multicolumn{1}{c|}{-} & 91.49 & 78.22 & 78.09 & 84.07 & 52.71 \\

\bottomrule
\end{tabular}%
\begin{tablenotes}
    \item[$\ast$] FLOPs are reported as MAC operations.
    \item[$\dagger$] Favorable batch size 100.
    \item[$\ddagger$] The C3D model used is different/smaller version from the original one \cite{Ji2013}.
\end{tablenotes}
\end{threeparttable}
}
\end{table*}

To evaluate the performance of the tool, four state of the art 3D CNN HAR models have been selected, Slowonly, R(2+1)D-18, R(2+1)D-34, and X3D (as shown in Table \ref{tab:models_characteristics}), alongside two FPGA platforms ZCU104 and ZCU102 to demonstrate the ability of the tool to target multiple 3D CNN with different workloads and network parameters on a variety of platforms. C3D model was also included to provide direct comparisons with existing works, the majority of which are hand-tuned, model-specific architectures and not toolflows. Vitis HLS and Vivado Design Suite (v21.2) were used, while the reported resource results are after place and route at 160 MHz clock frequency. The arithmetic precision used was 16-bit fixed point arithmetic with $Q8.8$ format. The accuracy of the HAR models is evaluated on the UCF-101, following the same strategy as prior studies\cite{Soomro2012UCF101:Wild,Tran2018}.

\begin{figure}
    \captionsetup{justification=centering}
    \centering
    \includegraphics[width=0.9\linewidth,keepaspectratio]{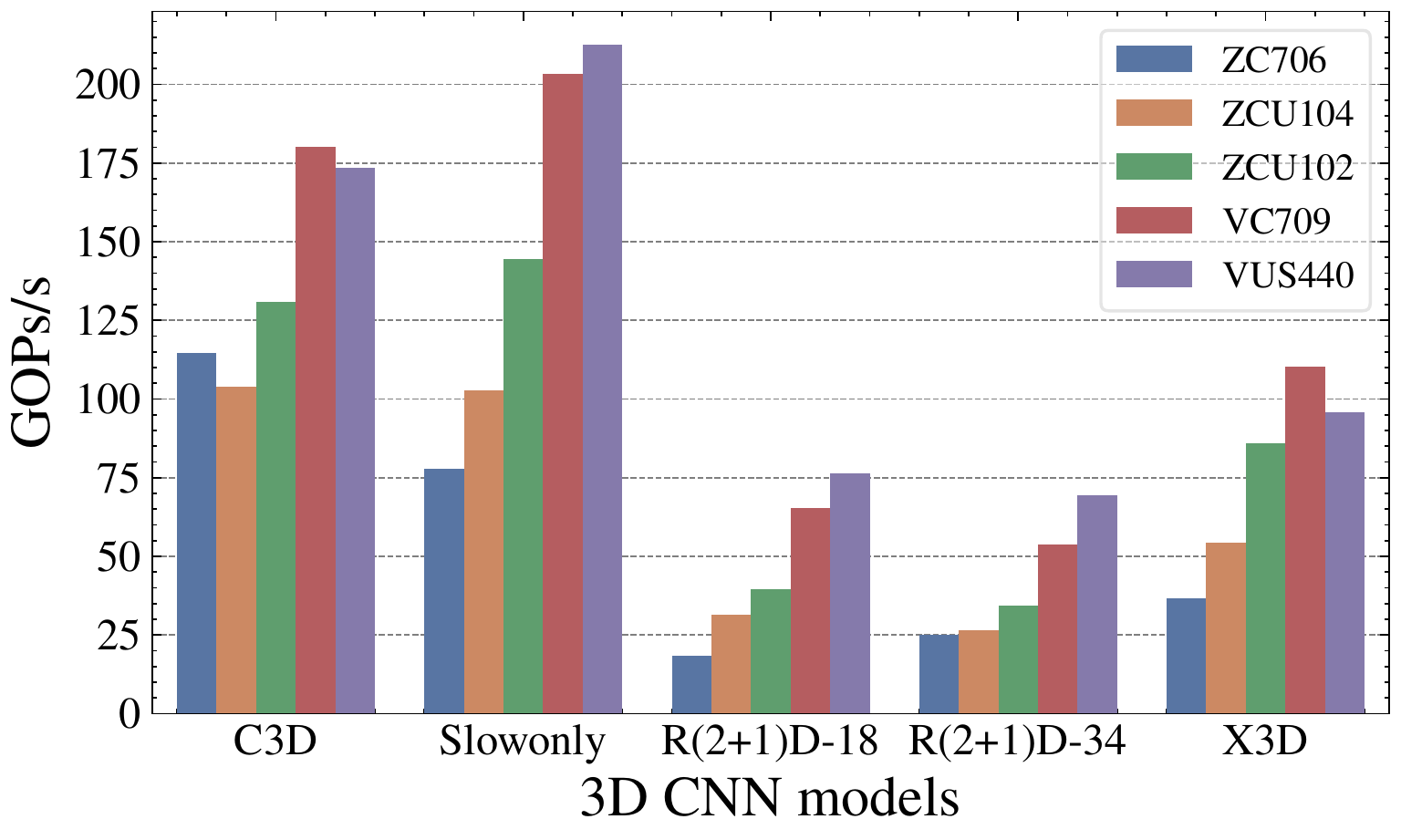}
    \caption{Throughput (GOPs/s) of fpgaHART-generated designs on 3D CNN HAR models delivering high-throughput results on a variety of FPGA devices}
    \label{fig:performance_devices}
\end{figure}
%

%

\subsection{Modeling Accuracy Evaluation}

To evaluate the quality of the performance predictor a series of experiments was conducted. Four partitions were chosen to cover the variety of produced graph structures (i.e. branch, sequential, multi-inputs, multi-outputs). The relative error was used to measure the difference between the predicted and actual latency. The relative errors for the four aforementioned types are 12.89\%, 5.03\%, 11.92\%, and 17.32\% respectively, giving a geometric mean relative error of 10.75\%\footnote{geometric mean was used as these structures can be found simultaneously in a single 3D CNN}. We found that the above errors are small enough to lead to meaningful design space exploration.

\subsection{Performance Comparison}

The fpgaHART has been evaluated on a number of different FPGA platforms, such as the ZC706, the ZCU102, the VC706, and the VUS440. Figure \ref{fig:performance_devices} displays the performance in GOPs/s (with a favourable batch size of 100) of the fpgaHART-generated designs for the 3D CNN models of Table \ref{tab:models_characteristics}, which details their unique characteristics, on a variety of FPGA devices. Such batch sizes are frequently encountered in practise when generating multiple views and clips over time and averaging them to improve the performance of the predictions. Even larger batch sizes may be required for multi-person HAR systems that evaluate each person's actions independently, as well as for large-scale systems that simultaneously analyse several videos.

The placement of fpgaHART in comparison to the rest of the existing works is outlined in Table \ref{tab:performance_comparison}, where the fpgaHART results are reported using ZCU102 as the FPGA platform. A conclusion readily apparent from Table \ref{tab:performance_comparison} is that fpgaHART is capable of delivering competitive performance on several 3D CNNs that have not been previously addressed and have a broad set of workloads and network parameters.

\begin{figure}[h]
    \captionsetup{justification=centering}
    \centering
    \includegraphics[width=0.9\linewidth,keepaspectratio]{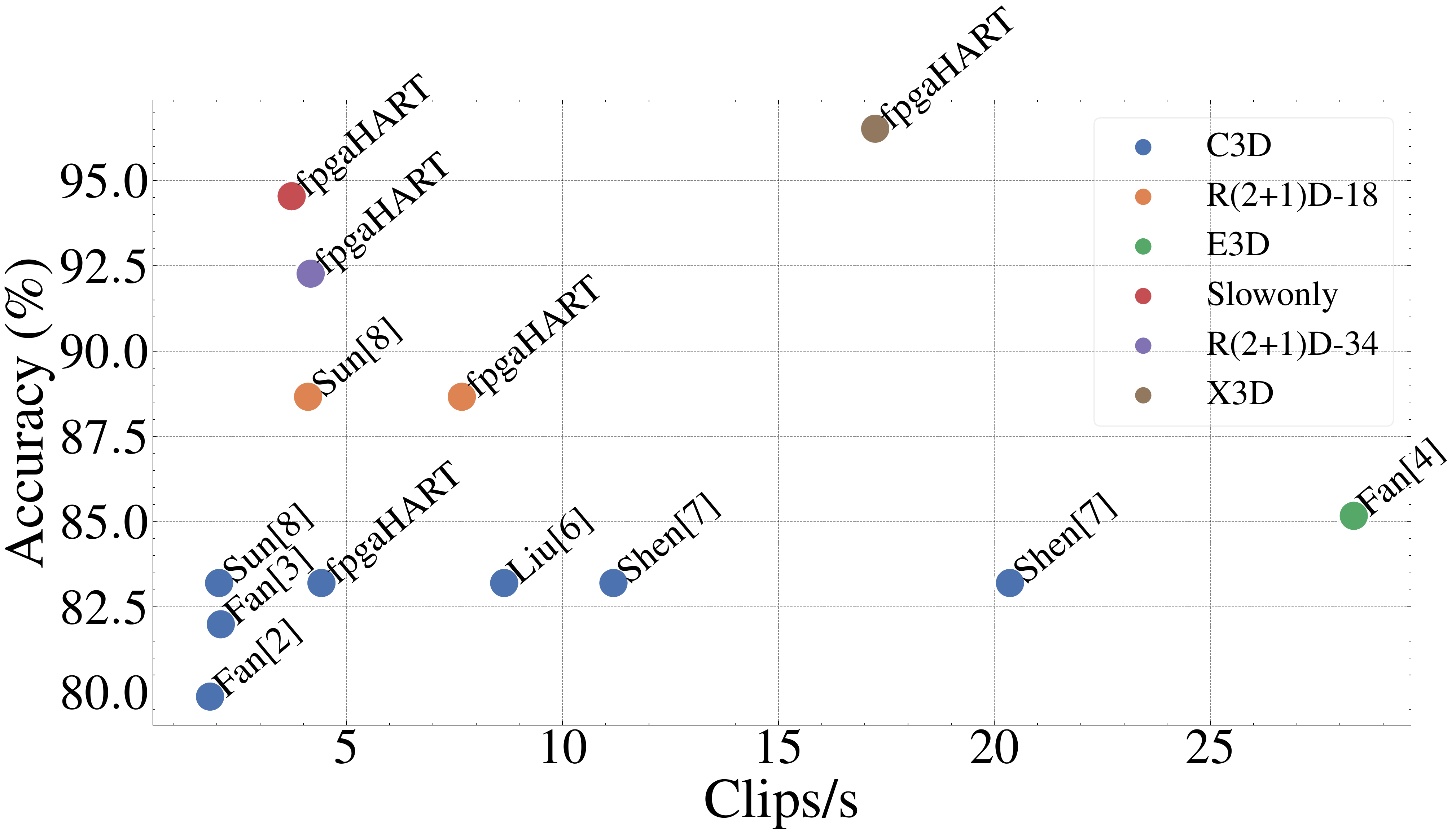}
    \caption{Pareto front on 3D CNNs: Clips/s over Accuracy. The fpgaHART results were taken using the VC709 FPGA platform, delivering solutions on the Pareto front.}
    \label{fig:performance_comparison}
\end{figure}

Figure \ref{fig:performance_comparison} presents the current state of the Pareto front expressed in terms of accuracy over throughput (clips/s), where the fpgaHART generated designs were derived targeting the VC709 FPGA platform. The results show that the fpgaHART models have pushed the Pareto front, delivering solutions with both high throughput and high accuracy, as shown in the graph.

Comparing the results on C3D (batch size 30 and targeting the ZCU102) to Nvidia RTX 3090, a server-grade GPU with 10496 CUDA cores and 1.7 GHz clock speed, the proposed architecture achieves a throughput of 4.42 clips/s compared to 281.87 clips/s that the GPU delivers. Yet, the proposed solution consumes only 26 W compared to the GPU's 298.6 W (excluding the CPU power consumption that a GPU system requires), offering 0.17 clips/s/watt compared to the GPU's 0.94 clips/s/watt.

%% file: Conclusions/conclusions.tex
\section{Conclusion} \label{conclusion}

This paper proposes an automated toolflow for the deployment and mapping of 3D CNN models for HAR onto FPGA devices. The proposed method employs SDF theory to describe and map 3D CNNs to hardware architectures. We demonstrate that the tool supports a pool of 3D CNNs for HAR on a variety of FPGA devices, while exhibiting comparable throughput performance to hand-tuned techniques. Future work may involve expanding the design space with additional SDFG transformations and improving the tool to support and provide latency-driven optimisation-focused designs.